%% file: low_f_noise_resub_with_bbl.tex
\documentclass[aps,prb,reprint]{revtex4-1}
\usepackage{CJK}
\usepackage{graphicx}
\usepackage{hyperref}

\begin{document}
\begin{CJK*}{UTF8}{gbsn}

\title{Low frequency noise peak near magnon emission energy in magnetic tunnel junctions}

\author{Liang Liu (刘亮)}
\author{Li Xiang (向黎)}
\author{ Huiqiang Guo (郭会强)}
\author{ Jian Wei (危健)}\email{weijian6791@pku.edu.cn}
\affiliation{International Center for Quantum Materials, School of Physics, Peking University, Beijing 100871, China}
\affiliation{Collaborative Innovation Center of Quantum Matter, Beijing, China}
\author{D. L. Li}
\author{Z. H. Yuan}
\author{J. F. Feng}\email{jiafengfeng@iphy.ac.cn}
\author{X. F. Han}
\affiliation{Beijing National Laboratory of Condensed Matter Physics, Institute of Physics, Chinese Academy of Sciences, Beijing 100190, China}
\author{J. M. D. Coey}
\affiliation{CRANN and School of Physics, Trinity College, Dublin 2, Ireland}

\date{\today}

\begin{abstract}
We report on the low frequency (LF) noise measurements in magnetic tunnel junctions (MTJs) below 4 K and at low bias, where the transport is strongly affected by scattering with magnons emitted by hot tunnelling electrons, as thermal activation of magnons from the environment is suppressed. For both CoFeB/MgO/CoFeB and CoFeB/AlO$_{x}$/CoFeB MTJs, enhanced LF noise is observed at bias voltage around magnon emission energy, forming a peak in the bias dependence of noise power spectra density, independent of magnetic configurations. The noise peak is much higher and broader for  unannealed AlO$_{x}$-based MTJ, and besides Lorentzian shape noise spectra in the frequency domain, random telegraph noise (RTN) is visible in the time traces.  During repeated measurements the noise peak reduces and the RTN becomes difficult to resolve, suggesting defects being annealed.  The Lorentzian shape noise spectra can be fitted with bias-dependent activation of RTN, with the attempt frequency in the MHz range, consistent with magnon dynamics. These findings suggest magnon-assisted activation of defects as the origin of the enhanced LF noise.
\end{abstract}
\maketitle
\end{CJK*}

Magnetic tunnel junctions (MTJs) have useful applications as magnetic sensors and memory elements. The tunneling barrier used is AlO$_{x}$ first, and then crystalline MgO since much enhanced magnetoresistance (MR) is observed due to spin dependent coherent tunneling.~\cite{Parkin2004nmat,*Yuasa2004nmat}
For application purposes, signal to noise ratio is important, and the low frequency (LF) noise has been intensively investigated for both AlO$_{x}$ based~\cite{Nowak1998jap,*Nowak1999apl,Ingvarsson1999jap,*Ingvarsson2000prl,Jiang2004prb} and MgO-based MTJs.~\cite{Guerrero2005apl,Gokce2006jap,Scola2007apl,Mazumdar2007apl,*Ozbay2009apl,*Stearrett2010apl,*Yu2011apl,*Stearrett2012prb,Zhong2011jap,Herranz2011apl,Cascales2013apl,Arakawa2012prb}
Field independent noise is usually considered of electrical origin, e.g., due to defect states or charge trapping; field dependent noise  is of magnetic origin, e.g., due to fluctuations of magnetic domains in the ferromagnetic electrodes. However, in spite of such categorization, the exact microscopic mechanisms are still not clear for both the 1/f and random telegraph noise (RTN). The influence of phonon emission on  LF noise has been considered for nano point contacts long time ago,~\cite{Akimenko1984,Kozub1993prb} and  for spin-valve type nanopillars magnon emission~\cite{Urazhdin2003prl,Kozub2007prb} was also considered. As the low bias transport of MTJs is strongly affected by scattering with magnons emitted by hot tunnelling electrons, here we study the possible influence of magnon emission on LF noise in MTJs.

To characterize the bias regime that magnon emission as well as other inelastic scattering starts to affect transport properties, second derivative of the current voltage characteristics is often obtained and it is called inelastic electron tunneling spectroscopy (IETS). In practice, from low to high bias, features in IETS are attributed to electron-electron interaction (EEI) near zero bias, magnon excitation in the interface between MgO and ferromagnetic electrodes around 20 mV, and phonon excitation of the MgO surface around 80 mV.~\cite{Han2001apl,*Bang2009jap,*Drewello2009prb,*Bernos2010prb,*Ma2011prb}  For MTJs the resistance in the parallel state (P) doesn't change much with bias or temperature, but changes a lot in the antiparallel state (AP), which is caused by enhanced emission of magnons in the AP state by hot tunnel electrons~\cite{Zhang1997prl} as characterized by a more pronounced magnon (`M') peak in IETS.

\begin{figure}
\includegraphics[width=9cm]{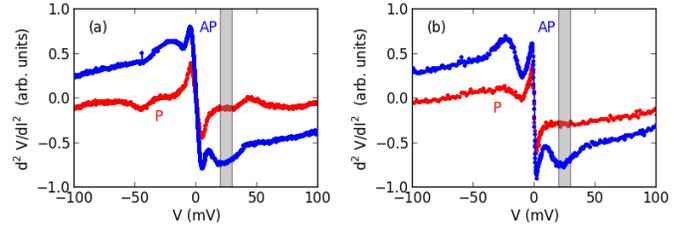}
\caption  {\small IETS in the P state (red line) and the AP state (blue line) at 3.6 K for MgO-based (a) and AlO$_{x}$-based (b) MTJs.  Magnon emission is illustrated by the shaded area around 20-30 mV. As the dip of $d^{2}V/d^{2}I$  corresponds to a peak in $d^{2}I/d^{2}V$, it is called `M' peak in the IETS literature. } 
\label{fig_dVdI}
\end{figure}

MgO-based MTJ stacks with the main structure of Ir$_{22}$Mn$_{78}$ (10), Co$_{90}$Fe$_{10}$ (2.5), Ru (0.9), Co$_{40}$Fe$_{40}$B$_{20}$ (3), MgO (2.5), Co$_{40}$Fe$_{40}$B$_{20}$ (3) (thickness in nanometers) were grown with a high vacuum Shamrock cluster deposition tool. The stack was then patterned into junctions with the rectangular shape of $5\times 10$ $\mu m^{2}$ using ultraviolet (UV) lithography and Ar ion beam etching, and annealed in an in-plane field of 8000 Oe at 350 $ ^{\circ}$C for half an hour to define the exchange bias of the antiferromagnetic IrMn layer and crystallize both the bottom and top CoFeB electrodes. For AlO$_{x}$-based MTJs of the same shape, the central layers are Co$_{40}$Fe$_{40}$B$_{20}$ (4), AlO$_{x}$ (1), Co$_{40}$Fe$_{40}$B$_{20}$ (4), and no annealing is conducted. 

First and second derivatives are measured with a digital lock-in amplifier with a DC bias circuit. Note that the dip of $d^{2}V/d^{2}I$  corresponds to a peak in $d^{2}I/d^{2}V$,~\cite{Holweg1992prb} and the so-called magnon `M' peak is clearly visible in Fig.~1 for two MTJs reported here. Magnetoresistance along the easy axis is shown in Fig.~2(f) and Fig.~3(d),  and the vertical dashed lines at -300 Oe and 500 Oe indicate AP and P states where IETS in Fig.~1 is measured. The sample is cooled in a cryogen-free dilution refrigerator equipped with a superconducting vector  magnet. The voltage fluctuations are AC coupled to two home-made amplifiers and then digitized with a data acquisition card. A cross correlation algorithm is used to average out the instrument noise when it is feasible. To measure the bias dependence of the LF noise, we use a digital voltage source decoupled with a home-made battery-powered optical decoupler circuit, and with a ballast resistor it is converted to a low noise current source.~\footnote{The purpose of the decoupling circuit is to decouple the ground of the voltage source from the ground of the battery-powered biasing circuit. See the datasheet of the optocoupler at http://www.ti.com/product/til300. A low pass filter is also added to attenuate the shot noise from the photodiode in the optocoupler.} For fixed current noise measurements, just battery and ballast resistors are used to avoid extra noises due to external circuit.

\begin{figure}
\includegraphics[width=9cm]{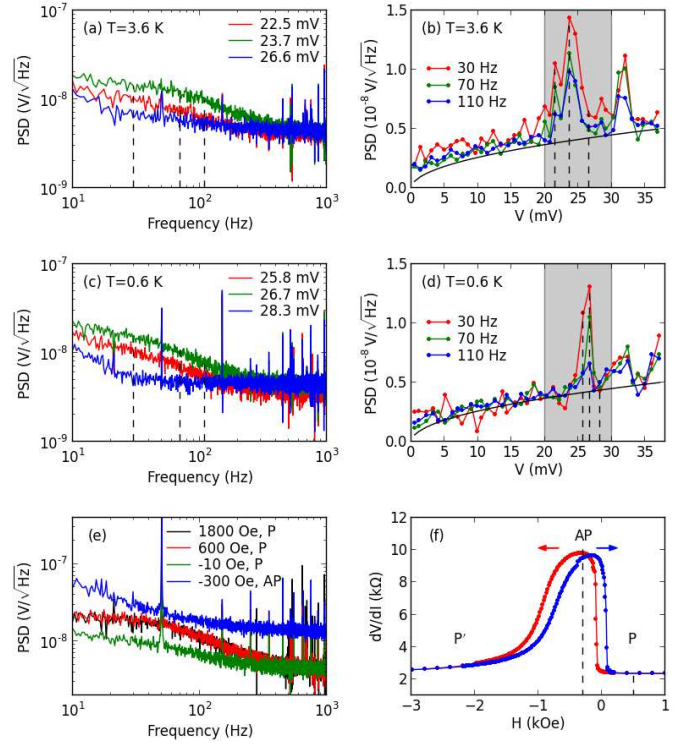}
\caption  {\small (color online) Noise PSD in the P state for a MgO-based MTJ at different bias voltage values around the magnon peak at 3.6 K (a) and 0.6 K (c), the noise plateau is visible. The dashed lines in (a) and (c) are the frequencies for which the bias dependence of PSD are presented in (b) and (d). The black lines in (b) and (d) show calculated full shot noise, and the shaded area marks the magnon peak. (e) Field independent voltage noise at a fixed bias current of 11 $\mu$A at 0.6 K. (f) Magnetoresistance with field ramping up (blue) and down (red). The vertical dashed lines at -300 Oe and 500 Oe indicate AP and P states where magnetoresistance and noise are measured.  There is a system noise which dominates below 20 Hz in (a) and (c), and absent when no extra circuit is used in (e).} 
\label{fig_PSD_MgO}
\end{figure}

Among three MgO-based MTJs, only one shows a small but clear Lorentzian shape plateau in the frequency domain, as shown in Fig.~2. For clarity, we present the voltage noise ($V_{N}$) power spectra density (PSD, in unit of $V/\sqrt{Hz}$) between 10 Hz and 1 kHz.  
The bias dependence of the noise PSD is a combination of a few peaks and the shot  noise background, with the latter frequency independent and proportional to $\sqrt{V}$  (for full shot noise $S_{I}\equiv I_{N}^{2}=2eI$, $V_{N}=I_{N}R_{AC}\propto\sqrt{V}$ when $R_{AC}\sim R_{DC}$), as denoted by the black lines in Fig.~2(b) and 2(d). 

The Lorentzian line shape is usually related to two level fluctuators (TLFs) and is described by~\cite{Machlup1954jap,Holweg1992prb} 
\[ 
S_{V}(f)=\frac{S_{0}\tau_{eff}}{1+(2\pi f \tau_{eff})^{2}},
 \]
where $S_{0}$ is the integrated power, $\tau_{eff}$ the effective time constant, $f$ the measurement frequency. For convenience we rewrite the equation as
\begin{equation}
 V_{N}(f)=\sqrt{S_{V}(f)}=\frac{V_{0}/\sqrt{f_{c}}}{\sqrt{1+(f/f_{c})^{2}}}=\frac{V_{0}\sqrt{f_{c}}}{\sqrt{f_{c}^{2}+f^{2}}},
\label{Eq_S_V}
\end{equation}
where $f_{c}=1/(2\pi\tau_{eff})$ is the characteristic frequency. We note that $S_{0}$ and $V_{0}$ determine the height of the flat top of the Lorentzian line shape when $f\ll f_{c}$.

The bias dependence of $V_{N}$ can be modelled in a simplified equivalent defect temperature scenario,~\cite{Holweg1992prb} i.e.,  $k_{B}T_{defect}= eV$, and with thermal activation 
\begin{equation}
\tau_{eff}(V)=\tau_{0}\exp(E/V),\quad f_{c}(V)=f_{0}\exp(-E/V),
\label{Eq_tau_eff}
\end{equation} 
where $E$ is the field-independent activation energy.  This model can also be modified to accommodate field dependence.~\cite{Ingvarsson2000prl,Arakawa2012prb}  
 With increasing bias $V$, $f_{c}(V)$ gets larger, the flat top $V_{0}/\sqrt{f_{c}(V)}$ becomes lower for $f\ll f_{c}$. When $f\sim f_{c}$, since $f_{c}(V)^{2}+f^{2}\geq 2f_{c}(V)f$ in Eq.~(\ref{Eq_S_V}), $V_{N}(V)\leq V_{0}/\sqrt{2f}$ and shows a maximum when $f = f_{c}(V)$, as is better illustrated in Fig.~3(b) for the AlO$_{x}$-based MTJ with peaks at different bias values for particular $f$. Here in Fig.~2b and 2d, the noise peaks for three different frequencies almost overlap together, which  suggests a weak bias dependence of $f_{c}$ according to Eq.~(\ref{Eq_tau_eff}) and this makes fitting difficult. Similar observation for nano point contact was ascribed to the electron phonon scattering,~\cite{Kozub1993prb,Akimenko1984}  magnon emission may be more relevant here. Additionally, the PSD peak width also shrinks from  about 4 mV at 3.6 K to about 2 mV at 0.6 K (peak at 18 K is much broader, not shown here), which suggests thermal smearing of the magnon emission threshold.

As shown in Fig.~2e, there is no field dependence of the noise plateau in the P state, and the plateau exists also in the AP state (not shown here). 
In addition, this small noise plateau is observed only in one direction of the applied bias, which is when the electrons flow from the pinned layer to the free layer. This may relate to the asymmetry of non-equilibrium magnon distribution as proposed before for phonons,~\cite{Kozub1993prb} i.e., as the defect is in one side of the barrier, only when there is magnon emission in this side can the fluctuators be activated.

When only a few TLFs dominate the spectra and there is no correlation between them, discrete levels should be visible in the time trace as it is called RTN. Previously RTN was often observed at particular field values where the resistance changes abruptly,~\cite{Ingvarsson1999jap,Ingvarsson2000prl,Ozbay2009apl} and was ascribed to magnetic \textit{after effect}.~\cite{Guo2009apl} Field dependent RTN was also observed for submicron MTJs,~\cite{Herranz2011apl,Zhong2011jap,Arakawa2012prb} and a modified heating model was assumed to interpret the field dependence. Field independent RTN in MTJ is observed in Ref.~[\onlinecite{Liu2009jap}] and charge trapping near or in the oxide barrier was suspected, which is most relevant here. We note that, even for electrical noise, magnon dynamics may still play a role in the microscopic process, as the magnon dispersion is only determined by the internal field of the FM electrodes,~\cite{Miao2006jap,Bang2009jap} which is due to exchange bias larger than the external field.~\cite{Wei2010prb}

\begin{figure}
\includegraphics[width=9cm]{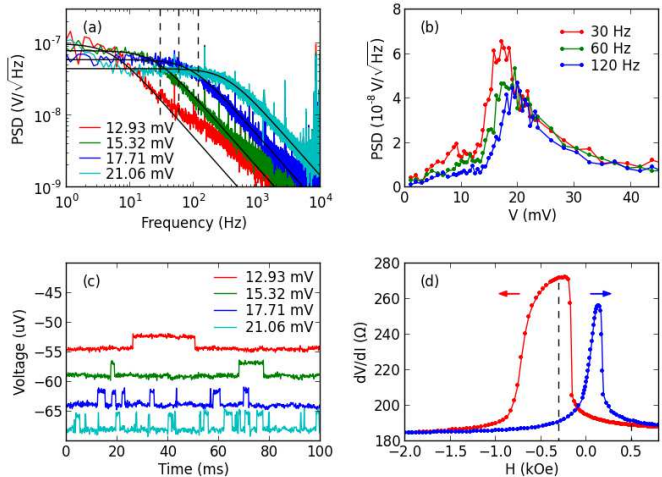}
\caption  {\small  (color online)  Noise PSD in the AP state for the AlO$_{x}$-based MTJ at different bias voltage values around the magnon peak at 3.6 K (a), and the bias dependence of PSD at different frequencies (b). The black lines in (a) are Lorentzian fits, and for the 12.93 mV data, there are two Lorentzian plateaus and only one fit is shown here (the overlap of TLFs is demonstrated in Fig.~4). The dashed lines in (a)  denote three frequencies for which the bias dependence of PSD are presented in (b). The shot noise is too small compared to LF noise so it is not shown in (b), but can be seen at relatively higher frequencies as shown in Fig.~5. (c) Reconstructed time traces showing the dominating  TLF in (a), with each trace shifted by a few $\mu$V for clarity. The voltage difference between two discrete states $\Delta V$ is about 2.5 $\mu$V. (d) Magnetoresistance  with field ramping up (blue) and down (red). } 
\label{fig_PSD_AlO}
\end{figure}

For one unannealed AlO$_{x}$-based MTJ, RTN is much larger and the switching is clearly visible  in the time traces, as shown in Fig.~3(a-c). We note that here the raw time trace data were digitized and Fourier transformed, after a low pass filtering the time trace is reconstructed by performing inverse Fourier transformation.  Shot noise is much smaller than LF noise so it is not plotted in Fig.~3(b). Since in this temperature and bias regime, TLFs are stable and correlations between TLFs are suppressed,  individual TLF can be well identified. The integrated spectra power $S_{0}$ for individual TLF is~\cite{Machlup1954jap,Holweg1992prb}
\begin{equation}
S_{0}=2\pi V_{0}^{2}=4(\Delta V)^{2}\frac{\tau_{eff}}{\tau_{1}+\tau_{2}}=(\Delta V)^{2}\frac{4\tau_{1}\tau_{2}}{(\tau_{1}+\tau_{2})^{2}},
\label{Eq_S_0}
\end{equation}
where $1/\tau_{eff}=1/\tau_{1} +1/\tau_{2}$, $\tau_{1}$ and $\tau_{2}$ are the mean times spending in the high and low level states, which can have their own activation forms similar to Eq.~(\ref{Eq_tau_eff}), and $\Delta V$ is the voltage difference of the two discrete levels. The bias dependence of the fitted parameters are presented in Fig.~4 in detail. 

When switching is rare, $\tau_{1}\ll \tau_{2}$, according to Eq.~(\ref{Eq_S_0}) $V_{0}\approx \sqrt{2/\pi}\sqrt{\tau_{1}/\tau_{2}}\Delta V\ll \Delta V$, which also follows an activation form as exemplified by the green symbols in Fig.~4(b) for the most visible TLF. In fact the activation energy and attempt frequency for $\tau_{1}$ and $\tau_{2}$ are 148, 247 mV, 0.33, 7.9 MHz respectively, and $\tau_{eff}$ is mostly dominated by $\tau_{1}$ at lower bias and by $\tau_{2}$ at higher bias, which explains the bending of $f_{c}$ in Fig.~4(a). Here although we follow the literature and discuss $\tau_{eff}$ or $f_{c}$ only, the activation energy and attempt frequency are in the same ballpark for $\tau_{1}$ and $\tau_{2}$. When the switching becomes more frequently at higher bias, $\tau_{1}\sim \tau_{2}$, $V_{0}$ approaches $\Delta V/\sqrt{2\pi}$, which is shown in Fig.~4(d). $\Delta V$ is close to  2.5  $\mu$V as shown in Fig.~3(c). In Fig.~4(b) and 4(d) $V_{0}$ approaches a 1  $\mu$V when bias is higher than 20 mV, which is consistent considering the factor $\sqrt{2\pi}\sim 2.5$.  In Fig.~4(f) the 30 Hz PSD is reconstructed with fitted $f_{c}$ and $V_0$, which overlaps very well with the measured PSD as in Fig.3(b), again suggest our fitting is consistent.

The fitted activation energy for $f_{c}$ of the most visible TLF is about 166 meV, comparable to 0.3$\times10^{-19}$ J that was found for nonmagnetic RTN of AlO$_{x}$-based MTJ head.~\cite{Liu2009jap} And the activation energy for the other three TLFs in Fig.~4(a) are 59, 111, and 358 meV respectively.  Note that for another effective temperature model~\cite{Ralls1988,*Ralls1989} used for RTN in Ref.~[\onlinecite{Arakawa2012prb}], $T_{defect}=T+\zeta V^{2}$ and the fitted activation energy there is 1.3$\times10^{-21}$ J, or 8 meV, even lower than the magnon peak. The difference seems to be whether the energy absorbed by the defect is due to equilibrium thermal energy ($\propto V^2$), or inequilibrium hot electron energy ($\propto eV$). The good fits with $eV$ dependence as shown in Fig.4 suggest that inequilibrium hot electron energy is a reasonable assumption here.

\begin{figure}
\includegraphics[width=9cm]{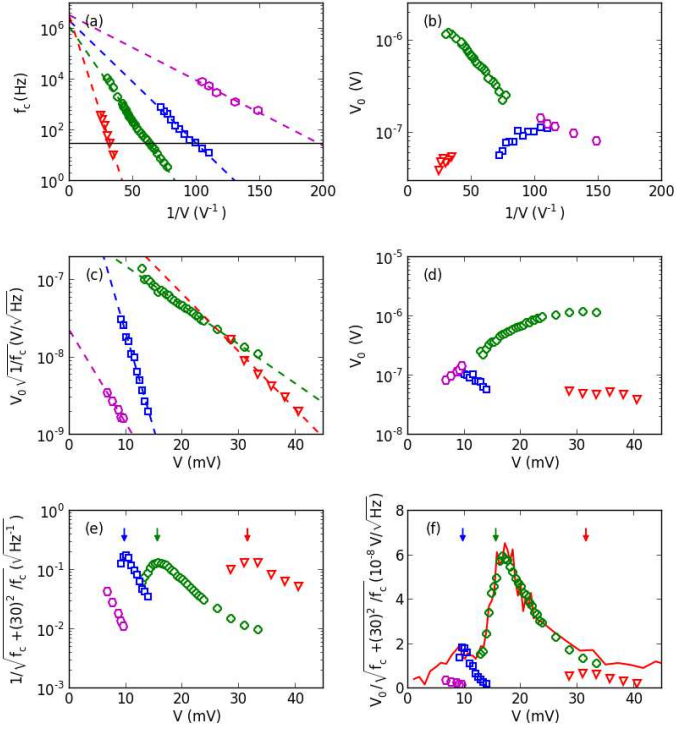}
\caption  {\small  (color online)  The bias dependence of $f_{c}$ (a) and height of the flat top $V_{0}/\sqrt{f_{c}}$ (c) are obtained with Lorentzian fits described by  Eq.~(\ref{Eq_S_V}) for AlO$_{x}$-based MTJ. Three of the four TLFs intercept with the $f$=30 Hz line and the cross points are indicated by arrows in (e) and (f). $V_{0}$ can be retrieved from $f_{c}$ and $V_{0}/\sqrt{f_{c}}$, and its bias dependence vs. $1/V$ and $V$ is shown in (b) and (d). There is no special feature between 20 to 30 mV as shown in (a) to (c), although $V_{0}$ does increase about an order of magnitude for the most visible TLF.   (e) The peak eventually shows up when the frequency dependent terms in Eq.~(\ref{Eq_S_V}) are combined, with $f$=30 Hz. (f) The reconstructed $V_N$($f$=30 Hz) are shown as the product of two factors in (d) and  in (e). The peaks overlap well with the red line which is the same curve in Fig.~3(b).} 
\label{fig_fits}
\end{figure}

For all four TLFs,  the attempt frequency  $f_{0}$ is around MHz, clearly lower than the typical GHz frequency for magnons, and THz frequency for phonons and atomic vibrations assumed in RTN of nanobridges.~\cite{Holweg1992prb,Ralls1988,*Ralls1989} The attempt frequency  was not explicitly discussed in previous measurements of RTN in MTJs.~\cite{Liu2009jap,Arakawa2012prb} However, for submicron MTJs it has been shown that the collective dynamics of many spins, i.e., the time it takes a domain wall segment to diffuse through a characteristic length, can be about 0.2 $\mu$s and this is just the microscopic attempt time for magnetization reversal et al.~\cite{Koch2000prl,Urazhdin2003prl} We believe that the same magnon dynamics in the FM electrode is responsible here for the activation of defects.

Two observations need to be emphasized. First,  enhanced LF noise is observed around 20 mV for both MgO and AlO$_{x}$-based MTJs, as well as for submicron MgO-based MTJ in Ref.~[\onlinecite{Arakawa2012prb}]. To check which factor of $V_{N}$ in Eq.~(\ref{Eq_S_V}) contributes to this enhancement, we plotted the two factors in Fig.~4(d) and 4(e) with $f$ set to 30 Hz. It is clear that while the peak position is due to the $f_{c}$ dependent factor, which is maximized when $f_{c}= f$ as shown in Fig.~4(e), the peak height is due to  $V_{0}$ in Fig.~4(d). In fact when there is a cross point with $f_{c}=30$ Hz in Fig.~4(a), there will be a peak, but the pronounced peak around 20 mV is due to the large $V_{0}$ for the most visible TLF, much larger than that of the other three. This can be related to magnon emission around this bias. Secondly, as shown in Fig.~3(c), $\Delta V$ for this particular TLF is about 2.5 $\mu$V and doesn't change much from 13 mV to 21 mV, inconsistent with a simple defect picture, where domain rotation or channel switching leads to fixed resistance change.~\cite{Holweg1992prb,Ralls1988,*Ralls1989} In fact almost fixed $\Delta V$ can also be extracted in previous reported cases,~\cite{Liu2009jap,Arakawa2012prb,Cascales2013apl}  although for measurements at high temperatures TLFs are  not stable and correlated with each other, thus the Lorentzian evolves to 1/f shape and detailed fitting with the activation model is not possible. How the fixed $\Delta V$ is related to magnon dynamics needs further investigation. 

\begin{figure}
\includegraphics[width=9cm]{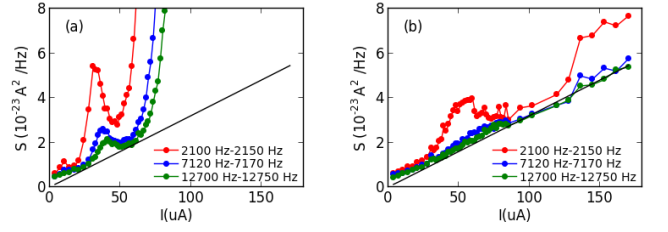}
\caption  {\small  (color online)  The bias dependence of PSD at higher frequencies for the AlO$_{x}$-based MTJ in early (a) and later (b) measurements. A 50 Hz band is used for averaging as it is much smaller than the frequencies here. The black lines show calculated full shot noise. Note that the same raw data are used to compute the PSD here in (a) and that in Fig.~3(b) while frequencies concerned are different. Also bias current dependence is presented here while in Fig.~2 and 3 voltage dependence is presented to emphasize magnon emission regime. } 
\label{fig_AlO_later}
\end{figure}

During repeated measurements the height of the noise peaks as shown in Fig.~3(b) reduces and their position changes slowly. Eventually shot noise background dominates again, similar to those in Fig.~2(b), as shown in Fig.~5.  This could be due to the fact that the AlO$_{x}$-based MTJs were not annealed during preparation and the defects can get annealed during the ramping of voltage or field, like those irreproducible TLFs reported early.~\cite{Holweg1992prb,Ralls1988,*Ralls1989,Gokce2006jap,Scola2007apl} Also since there is no crystallization of the CoFeB layer, the magnon emission at the interface may have a broad energy range. For the  MgO-based MTJ, we have repeated measurements in three cooldowns and the plateau is still observable, although $f_{c}$ slowly shifted to even lower frequencies in the last run and then it  can not be measured. 

We note that for both MgO and AlO$_{x}$-based MTJs, full shot noise is observed, even when there is a strong concurrent LF noise for the AlO$_{x}$-based MTJ (see Fig.~5). This indicates that the magnon dynamics doesn't affect the tunneling of electrons, and the tunnel barrier has no pin holes or trapped states, since otherwise the Fano factor (the ratio between the measured shot noise and full shot noise)  will be much  reduced from 1 when there is inelastic scattering or correlation effect in the barrier.~\cite{George2002,Guerrero2006} Recently it was revealed by IETS that defect states may exist in the barrier at energy higher than 150 mV.~\cite{Teixeira2011prl} Although this corroborates with the activation energy found here for the most visible RTN, but observation of full shot noise here suggests there is no sequential tunneling and no defect states inside the barrier. For comparison,  in the case of gold nanowire break junction the Fano factor is much reduced to ~0.02 and is affected by electron phonon scattering.~\cite{Kumar2012prl}

In summary, at low temperatures and low bias where the magnon emission dominates the transport properties of MTJs, low frequency noise peak in the bias dependence of power spectra density is observed. The noise peak for unannealed AlO$_{x}$-based MTJ is especially large, and RTN is visible in the time trace.  Detailed analyses of the RTN suggest defect activation by magnons dynamics. 

We thank X.-G. Zhang and Ryuichi Shindou for helpful discussions. Work at Peking University was supported by National Basic Research Program of China (973 Program) through Grant No. 2011CBA00106 and No. 2012CB927400. Work at IOP, CAS was supported by the State Key Project of Fundamental Research of Ministry of Science and Technology [MOST, No. 2010CB934401, 2014AA032904] and National Natural Science Foundation of China [NSFC, Grant No. 11434014].

\input{low_f_noise_resub.bbl}
\end{document}

%% file: low_f_noise_resub.bbl
%